\documentstyle[12pt]{article}
\begin{document}   
\centerline{\large{\bf {Light-neutrino mass hierarchies,}}} 
\centerline{\large{\bf {nuclear matrix elements,}}}
\centerline{\large{\bf {and the observability of neutrinoless $\beta \beta$ decay. }}} 
\centerline{O. Civitarese$^1$ and J. Suhonen$^2$} 
\centerline{\small{\it{$^{1}$ Dept. of Physics, Univ. of La Plata,
c.c. 67 (1900), La Plata, Argentina.}}} 
\centerline{\small{\it{$^{2}$ Dept. of Physics, Univ. of Jyv\"askyl\"a,
P.O.B 35, FIN-40014, Jyv\"askyl\"a, Finland.}}} 
\begin{abstract}
Results for neutrino flavor oscillations and neutrino mixing mechanisms, obtained from the  analysis 
of the Sudbury Neutrino Observatory (SNO), the SuperKamiokande (SK), 
CHOOZ, KamLAND and WMAP data, are used to calculate the effective neutrino mass 
$<m_\nu>$ relevant for the
neutrinoless double-beta decay ($0 \nu \beta \beta$). The observability of 
$0 \nu \beta \beta$ in the decay of $^{76}$Ge is discussed within different light-neutrino
mass hierarchies and by presenting a systematics on the available nuclear matrix elements. 
\end{abstract}



\newpage
\section{Introduction}
The knowledge about the properties of the neutrino has been
dramatically advanced by various large-scale experiments, 
as recently reported by the
SNO \cite{sno}, SK \cite{sk}, KamLAND \cite{kam}, CHOOZ \cite{chooz}, and WMAP \cite{wmap} collaborations.
These experimental evidences have confirmed the existence of 
neutrino flavor oscillations and have set stringent limits to the neutrino mass-mixing mechanisms.
A general overview of the latest experimental results
is given in recent review articles 
by Valle \cite{valle}, Bahcall et al.\cite{bahcall} and Elliot and Vogel
\cite{elliot}. Detailed discussions about the extracted 
values of the mixing angles,
mixing amplitudes, and mass differences can be found in 
Refs.\cite{bilenky,cheung,pascoli}. The implications of the
latest results on the physics of electroweak interactions and dark-matter studies
have been presented in Refs.\cite{ellis}-\cite{bhata}.
\footnote{Because of the large amount of publications in the field we focus our 
attention on the most recent ones, since most of the 
valuable previous literature
has been quoted in the papers which we have included in the present list of
references.}

In addition to the findings on neutrino flavor oscillations and the confirmation of some of the 
theoretically predicted possibilities for the mixing and 
enhancement of the oscillations in the presence of matter \cite{msw}, 
double-beta-decay experiments can
provide complementary information on the nature of
the neutrino and about its absolute mass scale \cite{smir}-\cite{fedor}.
This is a unique feature of the double-beta decay, which must be consistent with other scale-fixing
measurements, like the WMAP  measurements \cite{wmap}.
In the case of double-beta-decay measurements the knowledge about
relevant nuclear matrix elements is crucial, as it is crucial to know the correct 
neutrino-mass spectrum for the analysis of the other type of measurements.  
The implications of the results of the
solar, atmospheric, reactor and astrophysical neutrino experiments upon 
double-beta-decay experiments have been
stated already in several publications, see for instance 
\cite{kkps}-\cite{he}. To the wealth of parameters 
involved in the analysis, like CP-phases, mixing angles and 
masses, one should add the nuclear-structure degree of freedom needed to extract
the effective electron-neutrino mass \cite{report}-\cite{fedor}. 

At first glance, to physicists
who are less familiar with nuclear-structure analysis, it may seem an easy task to produce
the needed nuclear-structure information. 
Unfortunately it is not so because of several reasons: 

a) Double-beta-decay transitions 
take place in medium- and heavy-mass systems, where explicit shell-model calculations
are unfeasible, unless severely truncated valence spaces are used.

b) The sensitive part of the calculations depends
on the information about the structure of the states of the double-odd-mass nuclei.
These intermediate states play an essential role in 
the second-order  
transition matrix elements entering  the expression of the decay rate, and less is known about them, 
as compared with the relatively large amount of
information gathered about double-even-mass nuclei. 

c) In dealing with medium- and heavy-mass nuclei
one has to introduce approximations to obtain the participant wave functions and these approximations
are not unique, they vary from model to model.

d) To assign  a certain degree of significance to the already existing
theoretical results one has to define, first, what should be taken as the equivalent of 
the experimental confidence level, e.g. which models may be taken as references and what
would be the confidence level assigned to them depending upon the used approximations. 

In the past, all of these features have been referred to
as the {\it  uncertainties} in the nuclear matrix elements and roughly  estimated to be 
within factors of 2 to 3, with respect to some reference values. This aspect of the problem 
certainly deserves some attention, as we are going to discuss later on in this work,
since there turns out to be a gap between the range of masses extracted from double-beta-decay studies,
0.4 eV to 1.7 eV \cite{peter}, and those extracted from the other neutrino-related 
studies which yield upper limits of the 
order of 0.10 to 0.20 eV \cite{pierce} or even lower \cite{ellis}.
There is a clear discrepancy between both sets of results 
concerning the observability of neutrinoless double-beta ($0 \nu \beta \beta$) decay. This issue has become
a hot one, due to the recent claim \cite{kk} about the positive identification of neutrinoless-double-beta
decay signals in the decay of $^{76}$Ge (see also \cite{refutal,yuri}), 
from which a central value of the mass of the order of 0.39 eV
was extracted \cite{kk}. We think that these aspects must be considered from 
points of view of both the neutrino physics and the nuclear-structure physics. To achieve this goal,
in this work  we discuss the constraints set by the oscillation and mass parameters
on the effective neutrino mass measured in $0 \nu \beta \beta$-decay
transitions and compare these neutrino masses with the ones coming from
the analysis by using nuclear-structure information.
We start from the best-fit mass-mixing matrix presented in \cite{fit} and from
other estimates of the mixing matrix, i.e. the parametrization in terms of the
mixing angle of solar neutrinos, and the estimation based on a maximum-mixing 
scheme \cite{ma}.
  
In the first part of the paper we review the basic elements of the theory and 
discuss the structure of the adopted neutrino mass-mixing matrix. 
We discuss a way to extract light-neutrino masses ($m_i$) from the observed
mass differences and by combining them with the adopted neutrino mass-mixing matrix
we calculate the effective neutrino mass relevant for the $0 \nu \beta \beta$ decay.
In the second part of the paper we review the current nuclear structure information about
the $0 \nu \beta \beta$ decay, by presenting the up-to-date values
of the effective neutrino mass extracted from the adopted limits on the
half-lives. In doing so, we have considered the range of variation 
for the nuclear matrix elements, calculated within definite classes of models.
We have focused our attention to the case of the $0 \nu \beta \beta$ decay of
$^{76}$Ge. The analysis covers the calculated values of the nuclear 
matrix elements during the last fifteen years. 
This information is needed to estimate the plausibility of future
double-beta-decay experiments.
Finally, we discuss the observability of
the $0 \nu \beta \beta$ decay in the context of the present results.

\section{Formalism}
\subsection{Neutrino data}
Two- and three-generation analyses of neutrino data, provided by the solar
and atmospheric observations and by the range of mass differences explored
in reactor-based experiments, 
have been performed by several groups \cite{valle}-\cite{pascoli}.
The picture which emerges from these very detailed analyses of
neutrino-flavor oscillations
favours the Large Mixing Angle (LMA) solution of the Mikheyev-Smirnov-Wolsfenstein
(MSW) mechanism \cite{msw}. Recently, the KamLAND collaboration \cite{kam}
has confirmed the LMA solution and a crucial step towards the elucidation
of the neutrino-mass hierarchies was given by the results of WMAP
\cite{wmap,pierce}, which fix a stringent upper limit for the scale of neutrino masses.
A brief compilation of the adopted results is given in Table 1.
As shown in this table, the SNO data are
consistent with a value of 
the mass difference  
$\Delta m^2_{12}$ of the order of 10$^{-5}$ eV (solar-neutrino data), and another
independent scale
$\Delta m^2_{31} \approx \Delta m^2_{32}$, of the order of 10$^{-3}$ eV,
has been determined from the analysis of the atmospheric-neutrino data, which
is in the range of the sensitivity of the reactor-based measurements.
Because of the independence of the determined mass differences, 
the global picture is consistent with the existence of three active 
neutrino flavours. To these data, the information obtained by WMAP
is adding the value of the upper limit of the sum of the three
mass eigenvalues (light-neutrino masses only), which is of the order of 0.71 eV
\cite{wmap}.

To  calculate effective neutrino properties, like the effective electron-neutrino mass, 
$<m_\nu>$, one needs to know the neutrino-mixing matrix $U$ and the 
light-neutrino mass hierarchy $(m_1,m_2,m_3)$ \cite{smir}. 
The determination of the matrix elements of $U$
and the absolute values of the masses is the ultimate goal of any of
the models of the neutrino and it is, of course, a  matter of 
intensive effort.  
Out of the very rich, recently published list of articles dealing with 
the analysis of the SNO results, 
we have selected the results presented in 
the paper of Bandyopadhyay, Choubey, Goswami and Kar (BCGK) \cite{fit}, together with 
the expression of the mixing matrix in terms of the solar-neutrino data, and the
zeroth-order approximation of the mixing matrix assuming maximum mixing,
to perfom our calculations. Our choice is motivated by the fact that 
in the BCGK paper the best-fit value of $U$, with respect to the 
solar, atmospheric, and CHOOZ data, is written explicitly and the confidence level
of the results is well established.
  
The three-generation mixing matrix $U$ can be written as 
\begin{equation}
U=\left( \matrix{c_{13}c_{12}& s_{12}c_{13}&s_{13} \cr 
-s_{12}c_{23}-s_{23}s_{13}c_{12}&c_{23}c_{12}-s_{23}s_{13}s_{12}
 &s_{23}c_{13}\cr
 s_{23}s_{12}-s_{13}c_{23}c_{12}&-s_{23}c_{12}-s_{13}s_{12}c_{23}&
 c_{23}c_{13}\cr
  } \right) \;\;.
\end{equation}
This expression does not include CP violation.
By performing a three-generation $\chi^2$-analysis of the solar-neutrino and 
CHOOZ data, and by considering the mass differences 
$\Delta m^2_{12}=\Delta m^2_{\rm{solar}},
\Delta m^2_{31} \approx \Delta m^2_{32}=\Delta m^2_{\rm{atm}}$, 
the BCGK found that the best fit occurs in the LMA region with $tan^2 \theta_{13}
\approx 0$. This finding greatly simplifies the form of the mixing matrix $U$,
because it narrows the value of $U_{e3}$ down to a very small range around
$U_{e3} \approx 0$ \cite{ma,fit}. The best-fit 
form of $U$, reported in the BCGK paper, is
\begin{equation}
U=\left( \matrix{2\sqrt{\frac{2}{11}}& \sqrt{\frac{3}{11}}&0\cr 
-\sqrt{\frac{3}{22}}&\frac{2}{\sqrt{11}}
 &\frac{1}{\sqrt{2}}\cr
\sqrt{\frac{3}{22}} &-\frac{2}{\sqrt{11}} &
 \frac{1}{\sqrt{2}}\cr
  } \right) \;\;.
\end{equation}

The expression of the matrix $U$, considering $U_{e3}=0$ and exploiting the solar and atmospheric
mixing-angles data, reduces to (see Eq. (1))

\begin{equation}
U=\left( \matrix{c_{12}& s_{12}& 0 \cr 
-s_{12}c_{23}&c_{23}c_{12}
 &s_{23}\cr
 s_{23}s_{12}&-s_{23}c_{12}&
 c_{23}\cr
  } \right) \;\;.
\end{equation}

Finally, the matrix $U$, written in terms of maximum mixing 
($\rm{sin} \theta_{12}=\rm{cos} \theta_{23}=\frac{1}{\sqrt{2}}$),
is of the form

\begin{equation}
U=\left( \matrix{\frac{1}{\sqrt{2}}&\frac{1}{\sqrt{2}}&0\cr 
-\frac{1}{2}&\frac{1}{2}
 &\frac{1}{\sqrt{2}}\cr
\frac{1}{2} &-\frac{1}{2} &
 \frac{1}{\sqrt{2}}\cr
  } \right) \;\;.
\end{equation}

Only the first row is relevant for the electron-neutrino mass.
The next step consists of the definition of a neutrino mass hierarchy, that is the relative order of the
mass eigenvalues, which cannot be fixed only by the measured squared mass differences.

In order to estimate the possible range of the $m_i$, we
define the relative scales
\begin{equation}
m_1=fm_2 \;\;\;,
m_2=gm_3 \;\;\;
\end{equation}
for the normal mass hierarchy ($m_1\approx m_2 < m_3$), and 
\begin{equation}
m_1=fm_2 \;\;\;,
m_3=gm_1\;\;\;
\end{equation}
for the inverse ($m_1\approx m_2 > m_3$) and
degenerate ($m_3\approx m_2 \approx m_1$) mass hierarchies. 
To these factors we have added the information related to the 
scale of the mass eigenvalues, which is determined by the extreme value
\begin{equation}
m_0=\frac{\Omega_\nu}{3}
\end{equation}
where the value of $\Omega_\nu$ is taken from the WMAP data (see Table 1).
The factors $f$ and $g$ are determined in such a way that the
resulting masses $m_i(f,g)$ 
obey the observed mass differences, hereafter denoted
as $\Delta m^2$ ($\Delta m^2_{31} \approx \Delta m^2_{32}$)
and $\delta m^2$ ($\Delta m^2_{12}$). We are restricted to 
light-neutrino masses, as said before.
The numerical analysis was performed by assuming the above given scalings
and by finding the values of $(f,g)$ which are solutions of the equations
\begin{equation}
\frac{1}{1-g^2}-\frac{r}{1-f^2}=1\;\;\;
\end{equation}
for the normal hierarchy case and
\begin{equation}
\frac{r}{1-f^2}-\frac{1}{1-g^2}=r\;\;\;
\end{equation}
for the inverse and degenerate mass hierarchies. 
The use of the scale $m_0$ fixes the limiting values of $f$ and $g$ at 

\begin{eqnarray}
0 \leq f&\leq &\sqrt{1-\frac{\delta m^2}{g^2 m_0^2}} \;\;, \cr
0 \leq g&\leq &\sqrt{1-\frac{\Delta m^2}{m_0^2}}
\end{eqnarray}
for the normal mass hierarchy,
  
\begin{eqnarray}
0 < f&\leq&\frac{1}{\sqrt{1+\frac{\delta m^2}{m_0^2}}}  \;\;, \cr 
0 \leq g&\leq&\sqrt{1-\frac{\Delta m^2}{m_0^2}}
\end{eqnarray}
for the inverse mass hierarchy, and

\begin{eqnarray}
0 < f&\leq&\frac{1}{\sqrt{1+\frac{\delta m^2 g^2}{m_0^2}}}    \;\;,  \cr 
0 < g&\leq&\frac{1}{\sqrt{1+\frac{\Delta m^2}{m_0^2}}}
\end{eqnarray}
for the nearly degenerate masses.

In the above expressions the factor $r$ is given by the ratio between the solar and atmospheric
squared mass differences   
\begin{equation}
r=\frac {\delta m^2}{\Delta m^2}\;\;\;.
\end{equation}
Therefore, the variation of the parameters $f$ 
and $g$ is effectively restricted by the actual value of $r$ and $m_0$.

For each set of allowed values of $(f,g)$ and for each of the 
hierarchies considered we have calculated $m_i$. The effective neutrino mass $<m_\nu>$, relevant for the 
$0\nu \beta \beta$ decay, is given by \cite{report}-\cite{fedor}
\begin{equation}
{<m_\nu>}_{\pm}=\sum_{i=1}^3 m_i \lambda_i {\mid U_{ei}\mid}^2=m_1U_{e1}^2 \pm m_2U_{e2}^2\;\;\;,
\end{equation}
since for the adopted best fit $U_{e3} \approx 0$
\cite{fit}. We have consistently neglected CP violating phases and written $\lambda_i=\pm 1$,  
since the fit of \cite{fit} was performed under the assumption of CP conservation. In Table 2 we give, for 
each of the 
adopted forms of the mixing matrix U, the range of values of 
the calculated effective electron-neutrino masses.
These values correspond to the limiting values of $f$ and $g$, given in the previous equations (10)-(12).
As can be seen from this table, the largest value which one can obtain for $<m_\nu> $ 
is of the order of 0.24 eV, and the smallest one is of the order of 0.7 $\times 10^{-4}$ eV, both for the
degenerate mass hierarchy.
Notice that the larger value is of the order of the mass scale extracted 
from the results of WMAP and it will
certainly depend upon new results for $\Omega_\nu$. A value of $\Omega_\nu < 0.5 $ eV \cite{pierce} 
would then give a mass of the order of 0.16 eV, while the estimate  $\Omega_\nu < 0.18 $ eV \cite{ellis}  
will reduce it to the more stringent value of 0.06 eV.

\subsection{Nuclear matrix elements}

The implication of these results for 
$<m_\nu>$
upon the rates of $ 0\nu \beta \beta$ decay  is
easily seen if one writes \footnote{Only the mass
sector of the half-life will be considered in the following analysis. Complete expression of 
the half-life, including right-handed currents, can be found in \cite{report}.} the corresponding
 half-life, $t^{(0\nu)}_{1/2}$, as  
\begin{equation}
{\left(t^{(0\nu)}_{1/2}\right)}^{-1}=
{\left(\frac{<m_\nu>}{m_e}\right)}^{2}C_{mm}^{(0\nu)}\;\;\;,
\end{equation}
where the factor $C_{mm}^{(0\nu)}$
is defined as
\begin{equation}
C_{mm}^{(0\nu)}=G_1^{(0\nu)}{\left (M_{\rm GT}^{(0\nu)}(1-\chi_F)\right)}^2\;\;\;,
\end{equation}
in terms
of the nuclear matrix elements, $M_{\rm GT}^{(0\nu)}(1-\chi_F)$, and 
the phase-space factors, $G_1^{(0\nu)} $, entering the mass
term of the transition probability \cite{report}.

There are several aspects concerning Eq. (15) which are worth of mentioning:

a) In the event of a successful measurement of $ 0\nu \beta \beta$ decay and considering
the improving information emerging from  neutrino-related measurements, Eq. (15) may be viewed 
as a crucial test for nuclear models, since the calculated matrix elements reside in the factor 
$C_{mm}^{(0\nu)}$.  

b) If one assigns a certain confidence
level to nuclear-structure calculations, by fixing the value of $C_{mm}^{(0\nu)}$,
and takes the range of values of the effective neutrino mass
extracted from neutrino-related measurements, Eq. (15) 
may be viewed as a criterium for determining
the observability of $ 0\nu \beta \beta$ decay. 

c) In the event of a positive measurement of $ 0\nu \beta \beta$ decay 
and considering a reliable estimate of the nuclear
matrix elements, Eq.(15) may be viewed as a consistency equation 
for the effective neutrino mass seen in double-beta-decay
as compared with the one extracted from neutrino-related experiments.

Let us start with the discussion of the nuclear-structure related information, contained in
$C_{mm}^{(0\nu)}$.
The ultimate goal of nuclear-structure models is, in fact, the prediction of observables based on the
knowledge about nuclear wave functions down to the needed level of accuracy. In the case
of $ 0\nu \beta \beta$ decay studies, to achive this ultimate goal one needs to surpass several
requirements, some of which are purely technical and some of which are conceptual.
Among the technical barriers one has, of course, the unfeasibility of 
large-scale shell-model calculations, prohibited by hardware constraints. 
Among the conceptual requirements one has the 
realization that a prediction of a neutrinoless double-beta-decay rate should 
always be accompanied by other model predictions, like 
single-beta-decay, electromagnetic and particle-transfer transitions involving the
nuclei which participate
in the double-beta-decay transition under consideration. We stress the point that, in our experience, the
study should be conducted on the basis of a case-by-case analysis. 

Most of the current nuclear-structure approximations
are based on the proton-neutron quasiparticle random phase approximation (pnQRPA). This is a 
framework where 
proton-neutron correlations are treated as basic building blocks to describe the 
nuclear states which participate in
a double-beta-decay transition. The pnQRPA formalism is rather well known and it has
been discussed in a large number of publications during the last forty years. For
the sake of brevity we are not going to present it here again, rather we would like to
refer the interested reader to \cite{report} for details. In particular, the sensitivity
of the pnQRPA method to values of specific parameters of the interactions, like 
the sensitivity to the renormalization of the particle-particle (proton-neutron) coupling,
has been a matter of intensive studies. Again, we would like to refer
to \cite{report} for details concerning this point as well as concerning the large number of extensions
of the pnQRPA method, their successes and failures. Restricting ourselves to a very elementary 
theoretical background,
we can say that the standard procedure, applied in the literature to calculate 
the $0 \nu \beta \beta $-decay rate, involves three major components: 

a) The calculation of the spectrum of the intermediate double-odd-mass nucleus with  $(A,N\pm 1,Z \mp 1)$
nucleons. The pnQRPA is an approximate diagonalization in the one-particle-one-hole, 1p-1h, 
(or two-quasiparticle)
space and it includes the effects of 2p-2h ground-state correlations by means 
of the backward-going amplitudes.
Since the calculations are based on a quasiparticle mean field one forces
the breaking of certain symmetries, like the particle-number symmetry by the use of 
the BCS approximation, and the isospin symmetry, by the use of effective
proton and neutron single-particle states. The final results of the pnQRPA calculations 
will certainly be affected
by these symmetry-breaking effects induced by the way in which we handle the nuclear interactions.
Some attempts to cure for these effects have been implemented by means of enlarging the
representation space, including Pauli-principle-related blockings and by performing self-consistent 
approaches beyond the quasiparticle mean field. 
As said before, the list of various extensions of the standard 
pnQRPA is too long to be commented here in detail. A fairly complete list of references about
the set of extensions of the pnQRPA is given in \cite{report}-\cite{ejiri}.   
We will generally refer to these approximations
as pnQRPA-related ones. In this paper we shall show the results based on this family 
of approximations \footnote{We shall explicitly quote the sources from where the results have been taken
in order to avoid here a repetition of the details of each formalism, since the aim of the present
section is not to present a critical review of the theories but rather to show their results
to give an idea about the spread in the values of the relevant nuclear
matrix elements.}. In addition, we quote the results of the available shell-model calculations.

b) The calculation of the leptonic phase-space factors, as dictated by the second-order perturbative
treatment of the electroweak interaction. At the level of the minimal 
extension of the Standard Model (SM) lagrangian (mass sector only), 
these phase-space factors
can be easily calculated, and the values of them
should be rather universal causing no source of
discrepancies in the calculations, except for the adopted value of the axial-vector
coupling $g_A$. At the level of the two-nucleon mechanism this value is currently fixed at $g_A=1.254$
but for the medium-heavy and heavy nuclei an effective value of $g_A=1.0$ 
has also been used. In this work we adopt the conservative estimate of $g_A=1.254$.
Expressions for the phase-space factors, for theories beyond the minimal extension of the SM lagrangian,
i.e. for left-right and right-right couplings, have been listed exhaustively in the
literature (see e.g. \cite{report}-\cite{vergados}) and their values are 
well defined, too. In going beyond the two-nucleon mechanism
one has to consider, also, the momentum dependence of the operators, which will reflect upon
the structure of the phase-space factors. This is also true for the case of calculations where one is 
including p-electron wave effects and/or forbidden decays.
     
c) The calculation of the matrix elements of the relevant current operators which act upon the nucleons.
These operators are also well known and their multipole structures are derived from the expansion of
the electroweak current \cite{report}. In the present calculation we have considered the standard 
type of operators, without introducing further
momentum dependences in them, as originating from the electroweak decay at the quark level \cite{qrpa3}.

A compilation of 
the values of nuclear
matrix elements and phase-space factors can be found in \cite{report}.
The current information about the status of $0 \nu \beta \beta$ decays
is reported in \cite{zd02}-\cite{exp4}.

Tables 3 and 4 show the set of double-beta-decay systems where
experimental searches for signals of the
$0 \nu \beta \beta$ are conducted at present or planned for the next
generation of double-beta-decay experiments. The tables contain 
the experimental lower limits for the $0 \nu \beta \beta$
half-life \cite{you91}-\cite{kli86},
the full range of variation of the nuclear matrix elements, as 
contained in the factors  $C_{mm}^{(0\nu)}$ and as they
are predicted by different models \cite{report},
the values of the model-dependent factor \cite{report}    
\begin{equation}
F_N=t^{(0\nu)}_{1/2}C_{mm}^{(0\nu)}={(<m_\nu>/m_e)}^{-2} \;\;\;,
\end{equation}
the calculated phase-space factors $G_1^{(0\nu)}$, and
the extracted values of the effective neutrino mass.

In Table 4 only a sub-group of calculations are presented, namely the ones 
based on the plain spherical pnQRPA
approach of \cite{report} (third column). These results are compared with our present
calculations shown in the fourth column.

In the following, some brief details about the present pnQRPA calculations are given. 
They have been done by following the procedure outlined in \cite{report}.
The two-body nuclear interactions were constructed by using the G-matrix interaction of the Bonn type
including two to three major harmonic-oscillator shells around the proton and neutron Fermi surfaces.
The spherical Woods-Saxon potential was used to generate the single-particle energies and small adjustments
of these energies were done in the vicinity of the Fermi surfaces to reproduce the low-energy
quasiparticle spectra of the neighbouring odd-mass nuclei. 
Following the criteria which we have advanced above, the various parameters 
involved in the calculations have been fixed by reproducing the known data on single-beta-decay transitions
around the nuclei of interest for the double-beta-decay transitions which we are considering here. 
No further adjustement of the proton-neutron particle-particle coupling constant \cite{report}
is introduced once the known single-beta decay observables are reasonably reproduced.

As one can see, our present results are
in good agreement with the other pnQRPA calculations, except for $^{136}$Xe where our calculation gives
a larger matrix element than the other calculations. This deviation might occur due to the
semi-magic nature (the neutron shell is closed) of  $^{136}$Xe, forcing the transition from
the two-quasiparticle description to the particle-hole description.

If one compares the extracted neutrino masses of Table 3 with the ones given in the 
previous section, it becomes evident that the present generation of $0 \nu \beta \beta$ experiments
is rather insensitive to the effective neutrino mass coming from
the best fit of the solar+atmospheric+reactor data, except for the Heidelberg-Moscow
experiment if one takes the range of values ($<m_\nu>=$0.11 eV-0.56 eV) reported in \cite{kk}.
If one takes the heaviest possible effective mass 
$<m_\nu >\approx 0.24$ eV, which is favoured
 by the inverse and degenerate mass hierarchies (see Table 2),
one sees that it is outside the range of the present limits fixed by double-beta-decay experiments, 
with the possible exception of the decay of $^{76}$Ge, which just barely reaches this estimate.
 Naturally, the values of $<m_\nu >$
extracted from the experimental lower limits of $t^{(0\nu)}_{1/2}$ 
 are model dependent, since the connection between the half-lives and 
 the effective neutrino mass is given by the nuclear-model-dependent factors 
 $F_N$. As a reference value, for $<m_\nu >\approx 0.24$ eV  
one obtains $F_N=4.53 \times 10^{12}$ (see Eq.(17))
which is to be compared with the estimate (see Table 3, case of $^{76}$Ge ) 
$F_N= 3.55 \times 10^{11} \rightarrow 7.20 \times 10^{12}$, 
computed by assuming $t^{(0\nu)}_{1/2}> 2.5 \times 10^{25}$ yr \cite{yuri} and taking
into account the total span  of the calculated nuclear matrix elements.

With reference to the results shown in Table 4, the span in the effective neutrino 
mass is smaller (see the last column of Table 4) than when all the available model 
calculations are included (see the last column of Table 3). For the case of $^{76}$Ge 
the spherical pnQRPA gives a span of $<m_\nu>=$0.30 eV -0.43 eV in the effective mass. 
This means that to reach the neutrino-mass value resulting from the neutrino data, one definitely 
needs larger matrix elements than the ones produced thus far by the spherical pnQRPA model, and/or
longer half-lifes than the present measured limits. These observations will be discussed more in
the next section.

\subsection{pnQRPA Matrix elements for $^{76}$Ge} 

Table 5 shows the results of the matrix elements, corresponding to the mass sector of the neutrinoless
double-beta decay in $^{76}$Ge, calculated within the family of the pnQRPA-related models
\cite{staudt}-\cite{stoica}.
The standard spherical pnQRPA method gives results which are of 
the order of $C_{mm}^{(0)} \approx 5-8 \times 10^{-14}$
in units of yr$^{-1}$, with the exception of the result presented in \cite{stoica}, which 
yields to a magnitude of the order of $1.85 \times 10^{-14}$ yr$^{-1}$ , and the one of \cite{staudt}
where the pnQRPA value is $1.12 \times 10^{-13}$ yr$^{-1}$.
These factors translate into the ranges of 
the nuclear matrix elements \footnote{Notice that the results of \cite{staudt}, which are relevant 
for the analysis performed in \cite{kk}, are only 1.3 times larger than the average pnQRPA matrix element.}
and effective neutrino masses 
which were shown, previously, in Tables 3 and 4. The results
of the other, pnQRPA-related, approximations seem to be less stable and they deviate more 
from the central range of $C_{mm}^{(0)} \approx 5-8 \times 10^{-14}$
yr$^{-1}$. In analyzing the results of \cite{stoica} one can notice that the largest
value does not differ much from the standard pnQRPA value, although is has been
obtained by using a more involved approximation. By using the phase-space factors listed in
Table 4, we arrive at the central value for the matrix elements in the pnQRPA, namely
\begin{equation}
{{\rm{M}}_{\rm{GT}}^{(0\nu)}(1-{\chi}_{\rm{F}})}_{\rm{pnQRPA}}=3.65  \;\;\; .
\end{equation}
The corresponding value for the latest large-scale shell-model calculation \cite{etienne} is given by
\begin{equation}
{{\rm{M}}_{\rm{GT}}^{(0\nu)}(1-{\chi}_{\rm{F}})}_{\rm{shell-model}}=1.74 \;\;\; .
\end{equation}
Therefore, the latest shell-model results \cite{etienne} and the centroid of the pnQRPA results   
differ by a factor of the order of 2. In terms of the effective neutrino mass, using the
half-life limit $2.5 \times 10^{25}$yr, recommended in \cite{yuri}, these matrix elements
lead to
\begin{equation}
{<m_{\nu}>}_{\rm{pnQRPA}}=0.35 \;\; \rm{eV}   \;\; ,
\end{equation}
for the pnQRPA estimate, and 
\begin{equation}
{<m_{\nu}>}_{\rm{shell-model}}=0.74  \;\; \rm{eV}  \;\; ,
\end{equation}
for the shell-model estimate of the matrix element.
 It means that to go to masses of the order of 0.24 eV, as required by WMAP, one needs
larger nuclear matrix elements than the ones given by the pnQRPA or by the available shell-model results.
In fact, to reach the WMAP limit one would need  the value
\begin{equation}
{M_{\rm{GT}}^{(0\nu)}(1-{\chi}_{\rm{F}})}_{\rm{experimental}}=5.36,
\end{equation}
which is $\approx \sqrt{2}$ times larger than the reference pnQRPA value given in (18).
The largest matrix element listed in Table 5, coming from the VAMPIR approach \cite{vampir}, would 
yield to the value  ${{<m_{\nu}>}}_{\rm{VAMPIR}}=0.19$ eV, which just touches the
value  ${<m_{\nu}>}=0.24$ eV coming from the analysis of the neutrino-related data. 
However, it is appropriate
to point out here that the VAMPIR matrix element is to be considered unrealistically
large because in the calculations of \cite{vampir} no proton-neutron residual interaction
was included.

Finally, our present value
\begin{equation}
{{\rm{M}}_{\rm{GT}}^{(0\nu)}(1-{\chi}_{\rm{F}})}_{\rm{pnQRPA}}^{\rm{present}}=3.33
\end{equation}
(see Table 4) is consistent with the central value (18), and it yields an effective neutrino mass
\begin{equation}
{<m_{\nu}>}_{\rm{pnQRPA}}^{\rm{present}}=0.39 \;\; \rm{eV}  
\end{equation}   
if one takes for the half-life the lower limit recommended in \cite{yuri}, and
\begin{equation}
{<m_{\nu}>}_{\rm{pnQRPA}}^{\rm{present}}=0.50 \;\; \rm{eV}  
\end{equation}   
if one takes for the half-life the value $1.5 \times 10^{25}$yr given by the 
Heidelberg-Moscow collaboration \cite{kk}.

\section{Observability of the neutrinoless double beta decay}
 
To grasp an idea about the observability of the $0 \nu \beta \beta$ decay in other systems,
 we can compare the values of $F_N$, of Table 3, with the
 ones obtained by using an  effective neutrino mass
 of 0.24 eV, corresponding to  $F_N=4.53 \times 10^{12}$.

Figure 1 shows the comparison between the values of $F_N$ of Eq.(18),
listed in Table 3, and the values corresponding to the effective
neutrino masses $<m_\nu>=$ 0.24 eV and 0.39 eV. The interval between 
upper and lower values, for each case, represents the span of the
calculated nuclear matrix elements. For the case of $^{76}$Ge the
prominent upper value is given by the unrealisticaly large
nuclear matrix element of \cite{vampir}.

The results shown in Figure 1 indicate a departure with respect to the  
experimental limits by orders of magnitude, excluding the case of $^{76}$Ge which is closer but 
still outside of the range consistent with the solar+atmospheric+reator neutrino data.

 Thus the issue about the observability of the $0 \nu \beta \beta$ decay
 relies upon the estimates for the effective neutrino mass and upon
 the estimates of the  relevant nuclear matrix elements. 
 While in some cases
 the differences between the calculated matrix elements are within factors
 of the order of 3, in some other cases the differences are much larger. 
 It shows one
 of the essential features of the nuclear double-beta decay, namely that 
 case-by-case theoretical studies are needed instead of a 
 global one \cite{report}. 

\section{Conclusions}

To conclude, in this paper we have presented results on the effective
neutrino mass, as obtained from the best-fit mass-mixing matrix $U$ 
determined from the analysis of solar+atmospheric+reactor+satellite data,
and compared them with the values extracted from neutrinoless double-beta-decay
experiments. 

The effective electron-neutrino mass extracted from the neutrino-related experiments,
$<m_\nu> \approx 0.24$ eV, does not compare with the
central value of $<m_\nu> \approx 0.39$ eV , reported in \cite{kk} and obtained by using the
nuclear matrix elements calculated in \cite{staudt}. It does not compare, either, 
with the values given by the standard pnQRPA model, after taking into account the span
in the calculated matrix elements.   

To explain for the difference between the above results, we have compiled  a systematics
of the calculated nuclear matrix elements and performed additional pnQRPA calculations. 
In the case of $^{76}$Ge,
and if one adopts for the half-life  
the  upper limit of $ 2.5 \times 10^{25}$ yr suggested in \cite{yuri},
the nuclear matrix elements needed to yield the desired effective neutrino mass are 
larger than any of the known nuclear matrix elements calculated in the framework of 
the spherical pnQRPA. This is conclusion also holds for the available shell-model results.

The present knowledge of the
involved nuclear matrix elements shows that
the sensitivity of the $0 \nu \beta \beta$ experiments 
is still far from the estimate coming from neutrino oscillations.
However, the needed sensitivity is potentially achievable by 
the next generation of experiments.

\section*{Acknowledgments}

This work has been partially supported by the National Research Council 
(CONICET) of Argentina and by the
Academy of Finland under the Finnish Centre of Excellence Programme 2000-
2005 (Project No. 44875, Nuclear and Condensed Matter Programme at JYFL).
One of the authors (O.C) gratefully thanks the warm hospitality 
extended to him at the 
Department of Physics of the University of Jyv\"askyl\"a, Finland. 
The authors are grateful to Profs. J. Maalampi and A. Barabash for useful discussions.
\newpage
\begin{description}
\bibitem [1] {sno} Q. R. Ahmad et al. [SNO Collaboration],
Phys. Rev. Lett. 87 (2001) 071301; arXiv: nucl-exp/0204008,
nucl-exp/0204008.
\bibitem [2] {sk} S. Fukuda et al. [SuperKamiokande Collaboration],
Phys. Rev. Lett. 86 (2001) 5651.
\bibitem [3] {kam} K. Eguchi et al. [KamLAND Collaboration], 
Phys. Rev. Lett. 90 (2003) 021802.
\bibitem [4] {chooz}  M. Appollonio et al., 
Phys. Lett. B 466 (1999) 415.
\bibitem [5] {wmap} C. L. Bennet et al., arXiv:astro-ph/0302207. 
\bibitem [6] {valle}
J. W. F. Valle, arXiv:hep-ph/0205216, and references therein.
\bibitem [7] {bahcall}
J. N. Bahcall, M. C. Gonzalez-Garcia, C. Pe\~na-Garay, 
arXiv:hep-ph/0204314; arXiv:hep-ph/0204194.
\bibitem [8] {elliot}
S. R. Elliott, Petr Vogel, arXiv:hep-ph/0202264,
 submitted to Annu. Rev. Nucl. Part. Sci. 52 (2002), and references
 therein.
\bibitem [9] {bilenky}
S. M. Bilenky, D. Nicclo, S. T. Petcov, arXiv:hep-ph/0112216.
\bibitem [10] {cheung}
K. Cheung, arXiv:hep-ph/0302265.
\bibitem [11] {pascoli}
S. Pascoli, S. T. Petcov, W. Rodejohann, arXiv:hep-ph/0209059.
\bibitem [12] {ellis}
J. Ellis, arXiv:astro-ph:/0204059.
\bibitem [13] {minakata}
H. Minakata, H. Sugiyama, arXiv:hep-ph/0212240.
\bibitem [14] {bhata}
G. Bhattacharyya, H. P\"as, L. Song, T. Weiler,
arXiv:hep-ph/0302191. 
\bibitem [15] {msw} 
L. Wolfenstein, Phys. Rev. D 17 (1978) 2369; 
S. P. Mikheev, A. Y. Smirnov, Sov. J. Nucl. Phys. 42 (1985) 913.
\bibitem [16] {smir} A. Y. Smirnov, Czech J. Phys. 52 (2002) 439.
\bibitem [17] {report}
J. Suhonen, O. Civitarese, Phys. Rep. 300 (1998)123.
\bibitem [18]  {vergados}  
J. D. Vergados, Phys. Rep. 361 (2002) 1.
\bibitem [19]  {ejiri}     
H. Ejiri, Phys. Rep. 338 (2000) 265.
\bibitem [20]  {fedor}     
A. Faessler, F. \v Simkovic, J. Phys. G 24 (1998)  2139.
\bibitem [21] {kkps} H. V. Klapdor-Kleingrothhaus, H. P\"as, A. Y. Smirnov,
Phys. Rev. D 63 (2001) 073005.
\bibitem [22] {peter}
P. Vogel, Phys. Rev. D 66 (2002) 010001.
\bibitem [23] {ma}
E. Ma, arXiv:hep-ph/0303126.
\bibitem [24] {pierce} 
A. Pierce, H. Maruyama, arXiv:hep-ph/0302131.
\bibitem [25] {sarkar}
H. V. Klapdor-Kleingrothaus, U. Sarkar, 
Mod. Phys. Lett. 16 (2001) 2469.
\bibitem [26] {he}
H. J. He, D. A. Dicus, J. N. Ng, arXiv:hep-ph/0203237;
F. Feruglio, A. Strumia, F. Vissani, arXiv hep-ph/0201291.
\bibitem [27] {kk} H. V. Klapdor-Kleingrothaus, A. Dietz, H. L. Harney,
I. V. Krivosheina, Mod. Phys. Lett. A 16 (2001) 2409;
H. V. Klapdor-Kleingrothaus et al., Part. Nucl. Lett. 1 (2002) 57.
\bibitem [28] {refutal} C. E. Aalseth et al., arXiv:hep-ph/0202018, 
Mod. Phys. Lett. (in press).
\bibitem [29] {yuri}
Yu. G. Zdesenko, F. A. Danevich, V. I. Tretyak,
Phys. Lett.  B 546 (2002) 206.
 \bibitem [30] {fit}
A. Bandyopadhyay, S. Choubey, S. Goswami, K. Kar, arXiv:hep-ph/0110307.
\bibitem [31] {qrpa3} 
J. Suhonen, O. Civitarese, A. Faessler, Nucl. Phys.  A 543 (1992) 645.
\bibitem [32] {zd02} Y. Zdesenko, Rev. Mod. Phys. 74 (2002) 663.
\bibitem [33] {expbb1}
V. I. Tretyak, Yu. G. Zdesenko, At. Nucl. Data Tables
61 (1995) 43.
\bibitem [34] {expbb2}
F. A. Danevich et al., Phys. At. Nucl. 59 (1996) 1 .
\bibitem [35] {expbb3}
V. I. Tretyak, Yu. G. Zdesenko, At. Nucl. Data Tables
80 (2002) 83.
\bibitem [36] {exp4} A. S. Barabash, Czech J. Phys. 52 (2002) 567.
\bibitem [37] {you91}
Ke You et al., Phys. Lett. B 265 (1991) 53.
\bibitem [38] {bau99}
L. Baudis et al., Phys. Rev. Lett. 83 (1999) 41.
\bibitem [39] {ell92}
S. R. Elliot et al., Phys. Rev. C 46 (1992) 1535.
\bibitem [40] {arn99}
R. Arnold et al., Nucl. Phys. A 658 (1999) 299.
\bibitem [41] {eji01}
H. Ejiri at al., Phys. Rev. C 63 (2001) 065501.
\bibitem [42] {dan00}
F. A. Danievich et al., Phys. Rev. C 62 (2000) 045501.
\bibitem [43] {kal52}
M. I. Kalkstein, W. F. Libby, Phys. Rev.  85 (1952) 368.
\bibitem [44] {ale00}
A. Alessandrello et al., Phys. Lett. B 486 3319 (2000).
\bibitem [45] {lue98}
R. Luescher et al., Phys. Lett. B 434 (1998) 407.
\bibitem [46] {kli86}
A. A. Klimenko, A. A. Pomansky, A. A. Smolnikov, Nucl. Instr. and  Meth.
B 17 (1986) 445.
\bibitem [47] {staudt}
A. Staudt, K. Muto, H. V. Klapdor-Kleingrothaus, Europhys. Lett.
 13 (1990) 31.
\bibitem [48] {qrpa1}
K. Muto, E. Bender, H. V. Klapdor-Kleingrothaus, Z. Phys. A 334 (1989) 187.
\bibitem [49] {qrpa4}
G. Pantis, F. \v Simkovic, J. D. Vergados, A. Faessler,  Phys. Rev. C 53 (1996) 695.
\bibitem [50] {qrpa5}
T. Tomoda, Rep. Prog. Phys. 54 (1991) 53.
\bibitem [51] {qrpa6}
C. Barbero, F. Krmpotic, A. Mariano, D. Tadic,  Nucl. Phys.  A 650 (1999) 485.
\bibitem [52] {qrpa7}
F. \v Simkovic, G. Pantis, J. D. Vergados, A. Faessler,  Phys. Rev. C 60 (1999) 055502.
\bibitem [53] {qrpa8} 
A. Bobyk, W. A. Kaminski, F. \v Simkovic, Phys. Rev. C 63 (2001) 051301(R).
\bibitem [54] {stoica}
S. Stoica, H. V. Klapdor-Kleingrothaus,  Nucl. Phys. A 694 (2001) 269.
\bibitem [55] {vampir} 
T. Tomoda, A. Faessler, K. W. Schmid, F. Gr\"ummer, Nucl. Phys. A 452 (1986) 591.
\bibitem [56] {wick} 
W. C. Haxton, G. F. Stephenson Jr., Prog. Part. Nucl. Phys. 12 (1984) 409.
\bibitem [57] {etienne} 
E. Caurier, F. Nowacki, A. Poves, J. Retamosa, Phys. Rev. Lett. 77  (1996) 1954.
\bibitem [58] {aunola98}
M. Aunola, J. Suhonen, Nucl. Phys. A 643 (1998)  207.
\end{description}
\newpage

Figure Captions

Figure 1: Factors $F_N$, of Eq. (18), for each of the systems of Table 3.
          The lines are drawn to guide the eye. The interval between the upper and
          lower lines, for each case, represents the span of the 
          calculated nuclear matrix elements.
          The results corresponding to $<m_\nu>=$ 0.24 eV and
          $<m_\nu>=$ 0.39 eV are shown as
          horizontal lines.

\newpage
\begin{table}
\caption{Current limits on neutrino-mass differences.
The values listed are a compilation of the results from the 
SNO \cite{sno}, SK \cite{sk}, KamLAND \cite{kam} and WMAP \cite{wmap}.
\label{table1}}
\begin{tabular}{lc}
\hline
 ${\delta m^2}_{12}={\delta m^2}_{\rm solar}$& 
$5 \times 10^{-5} {\rm{eV}}^2 \rightarrow 1.1 \times 10^{-4} {\rm{eV}}^2$ \\
 ${\delta m^2}_{23}={\delta m^2}_{\rm atm}$& $10^{-3} {\rm{eV}}^2 
\rightarrow 5 \times 10^{-3} {\rm{eV}}^2$ \\
 ${\sin}^2{2\theta_{\rm solar}}$& $\approx$ 0.86 \\
 ${\sin}^2{2\theta_{\rm atm}}$&$\approx$ 1.0 \\
 $\Omega_\nu$ & $ < 0.71$ eV \\
\hline
\end{tabular}
\end{table}
\begin{table}
\caption{ Calculated effective electron-neutrino masses 
$ {<m_\nu>}_{\pm}$. Indicated in the table are
the mass hierarchy and the adopted mixing matrix. 
The values are given in units of eV. The results listed as {\it{extreme}}
 have been obtained by using the
extreme upper values of $f$ and $g$ of Eqs.(10)-(12).
The adopted values for the mass differences are   
${\delta m^2}_{12}= 7.1 \times 10^{-5} {\rm{eV}}^2$, 
 ${\delta m^2}_{23}= 2.7  \times 10^{-3} {\rm{eV}}^2$, and
$m_0$=0.24 eV. The mixing matrix U(a) is taken from the best fit of \cite{fit}, U(b)
is based on the largest values of the solar and atmospheric mixing angles, and U(c)
assumes maximum mixing.
\label{table2}}
\begin{tabular}{lccccc}
\hline
Hierarchy& & $<m_\nu>$ & U(a) & U(b)& U(c)\\ 
\hline
Normal &($m_1=0$) & $ {<m_\nu>}_{-} $ & -0.010& -0.012 & -0.019 \\
       &          & ${<m_\nu>}_{+}$ & 0.011 & 0.012 & 0.019 \\
       &(\it{extreme})& $ {<m_\nu>}_{-} $ & 0.105& 0.086 & -0.769 $\times 10^{-4}$ \\
       &              & ${<m_\nu>}_{+}$ & 0.231 & 0.231 & 0.231 \\
Inverse  &($m_3=0$)  & $ {<m_\nu>}_{-} $ & 0.105     & 0.087 & -0.153 $\times 10^{-2}$ \\
         &        & ${<m_\nu>}_{+}$ & 0.234& 0.235 & 0.235 \\
  &(\it{extreme})& $ {<m_\nu>}_{-} $ & 0.108& 0.088 & -0.749 $\times 10^{-4}$ \\
          &           & ${<m_\nu>}_{+}$ & 0.237 & 0.237 & 0.237 \\
Degenerate & (\it{extreme})     & $ {<m_\nu>}_{-} $ & 0.107& 0.088 & -0.715 $\times 10^{-4}$ \\
           &      & ${<m_\nu>}_{+}$     & 0.237& 0.237 & 0.237 \\
\hline 
\end{tabular}
\end{table}
\newpage
\begin{table}
\caption{$0 \nu \beta \beta$: Model-dependent estimates 
and experimental limits. 
 The double-beta-decay systems are
 given in the first column. The factors $C_{mm}^{(0 \nu)}$  
 are given in units of yr$^{-1}$ and their values are shown within
 the intervals predicted by different nuclear-structure models,
 like the Shell Model (a),
 the Quasiparticle Random Phase Approximation (b), 
 the pseudo SU(3) model (c), and various
 other models (d). The value $g_A=1.254$ is used.  
 The quantities $t_{1/2}^{(0 \nu)}$ are the 
 experimental lower limits of the half-lives, in units of years. The corresponding  
 references are quoted in brackets.
 The factor $F_N$ (lower limit) is shown in the fourth column and the values
  are given within the intervals provided by the factors
  $C_{mm}^{(0 \nu)}$. The last column shows the range of 
  variation of the extracted effective neutrino mass (upper limits)
  in units of eV. The coefficients  $C_{mm}^{(0 \nu)}$ are taken from \cite{report}, except for
the case of $^{124}$Sn \cite{aunola98}. 
\label{table3}}
\begin{tabular}{lcccc}
\hline
 System & $C_{mm}^{(0 \nu)}$ & $t_{1/2}^{(0 \nu)}$ & 
 $F_N$ & $ <m_\nu> $  \\
\hline
 $^{48}$Ca & (1.55-4.91) 10$^{-14}$ (a) & 9.5 10$^{21}$ \cite{you91} &
 (1.47-4.66) 10$^{8}$ & (23.7-42.1) \\
  & (9.35-363) 10$^{-15}$ (b) &   &
  (8.88-345) 10$^{7}$ & (8.70-54.2) \\
 $^{76}$Ge & (1.42-28.8) 10$^{-14}$ (d) & 2.5 10$^{25}$ \cite{yuri}  &
 (3.55-72.0) 10$^{11}$ & (0.19-0.86) \\
 $^{82}$Se & (9.38-43.3) 10$^{-14}$ (d) & 2.7
 10$^{22}$ \cite{ell92} & (2.53-11.7) 10$^{9}$ & (4.73-10.2) \\
 $^{96}$Zr & (9.48-428)10$^{-15}$ (b) & 1.0 10$^{21}$ \cite{arn99} &
 (9.48-428) 10$^{6}$ & (24.7-166) \\
 $^{100}$Mo & (0.07-2490) 10$^{-15}$ (b) & 5.5 10$^{22}$ \cite{eji01} & 
 (0.38-13700) 10$^{7}$ & (1.38-262) \\
 $^{116}$Cd & (5.57-66.1) 10$^{-14}$ (b) & 7.0 10$^{22}$ \cite{dan00} &
 (3.90-46.3) 10$^{9}$ & (2.37-8.18) \\
 $^{124}$Sn & (2.29-5.70) 10$^{-13}$ (b) & 2.4 10$^{17}$ \cite{kal52} &
 (5.50-13.7) 10$^{4}$ & (1.38-2.18) 10$^{3}$ \\
 $^{128}$Te & (1.71-33.6) 10$^{-15}$ (b) & 8.6 10$^{22}$ \cite{ale00} &
 (1.47-28.9) 10$^{8}$ & (9.51-42.1) \\
 $^{130}$Te & (1.24-5.34) 10$^{-13}$ (b) & 1.4 10$^{23}$ \cite{ale00} &
 (1.74-7.48) 10$^{10}$ & (1.87-3.87) \\
 $^{136}$Xe & (2.48-15.7) 10$^{-14}$ (a,b) & 4.4 10$^{23}$ \cite{lue98} &
 (1.09-6.91) 10$^{10}$ & (1.94-4.89) \\
 $^{150}$Nd & (4.78-77.4) 10$^{-13}$ (b,c) & 1.7 10$^{21}$ \cite{kli86} &
 (8.13-132) 10$^{8}$ & (4.45-17.9) \\
\hline
\end{tabular}
\end{table}
\newpage
\begin{table}
\caption{Calculated phase-space factors $G_1^{(0\nu)}$ and calculated nuclear matrix 
elements, using the formalism of the spherical pnQRPA, 
for some of the double-beta emitters included in Table 3. The phase space factors
are given in units  of  yr$^{-1}$ and the dimensionless matrix elements are scaled by the 
nuclear radius \cite{report}. The third column, indicated as N.M.E.,
gives the extreme values of the nuclear matrix elements 
$\rm{M_{GT}^{0 \nu}(1-\chi_F})$ reported in the literature (see 
the captions to Table 3), and the fourth column, indicated  as  N.M.E. (this work), 
gives the results of the present calculations for $\rm{M_{GT}^{0 \nu}(1-\chi_F})$.
The last column shows the range of values for the effective
neutrino mass, in units of eV, extracted from the results given in the third and fourth columns.
\label{table4}}
\begin{tabular}{lcccc}
\hline
 System & $ G_1^{(0\nu)} \times 10^{14}$ & N.M.E. & 
 N.M.E.(this work)  & $<m_\nu>$ \\
\hline
 $^{48}$Ca & 6.43  & 1.08-2.38   &           &  8.70-19.0 \\
 $^{76}$Ge & 0.63   & 2.98-4.33  &  3.33     & 0.30-0.43 \\
 $^{82}$Se &  2.73   & 2.53-3.98 &  3.44     & 4.73-7.44\\
 $^{96}$Zr &  5.70  &  2.74      &      3.55 &19.1-24.7   \\
 $^{100}$Mo & 11.30   &  0.77-4.67 & 2.97    &1.38-8.42  \\
 $^{116}$Cd & 4.68  &  1.09-3.46  & 3.75     &2.37-8.18 \\
 $^{128}$Te & 0.16  & 2.51-4.58 &            &9.51-17.4      \\
 $^{130}$Te & 4.14  &   2.10-3.59 & 3.49     &1.87-3.20    \\
 $^{136}$Xe & 4.37   &  1.61-1.90 & 4.64     &0.79-2.29    \\
\hline
\end{tabular}
\end{table}
\newpage
\begin{table}
\caption{Calculated nuclear matrix elements for the case of $^{76}$Ge. The values $C_{mm}^{(0)}$
are given in units of yr$^{-1}$. The adopted value for the half-life is the
value recommended in \cite{yuri}, $t_{1/2}^{(0 \nu)}=2.5 \times 10^{25}$yr. Indicated 
in the table are the models used to calculate the nuclear matrix elements, 
which are taken from the references quoted in the last row of the table.
The abreviations stand for the proton-neutron quasiparticle random-phase approximation (pnQRPA),
particle-number-projected pnQRPA (pnQRPA (proj.)), proton-neutron pairing pnQRPA (pnQRPA+pn pairing),
the renormalized  pnQRPA (RQRPA),
the second pnQRPA (SQRPA),
the self-consistent renormalized pnQRPA (SCRQRPA),
the fully renormalized pnQRPA (full-RQRPA), and the variation after projection
mean-field approach (VAMPIR). The model assumptions underlying these theories are presented in the quoted
references.  
 \label{table5}}
\begin{tabular}{lccc}
\hline
$C_{mm}^{(0)} $ & $F_N \times 10^{-12}$ & Theory & Reference \\
\hline
$1.12 \times 10^{-13}$ & 2.80  & pnQRPA & \cite{staudt,qrpa1}  \\
$6.97 \times 10^{-14}$ & 1.74  & pnQRPA & \cite{qrpa3}   \\
$7.51 \times 10^{-14}$ & 1.88  & pnQRPA (proj.) & \cite{qrpa3}   \\
$7.33 \times 10^{-14}$ & 1.83 & pnQRPA &  \cite{qrpa4}  \\ 
$1.42 \times 10^{-14}$ & 0.35  & pnQRPA+ pn pairing & \cite{qrpa4}   \\
 $1.18 \times 10^{-13}$ & 2.95  & pnQRPA & \cite{qrpa5}  \\
$8.27 \times 10^{-14}$ & 2.07 & pnQRPA & \cite{qrpa6}   \\
 $2.11 \times 10^{-13}$ & 5.27  & RQRPA &  \cite{qrpa7}   \\
 $6.19 \times 10^{-14}$ & 1.55  & RQRPA+ q-dep. operators &  \cite{qrpa7}  \\
 $1.8-2.2 \times 10^{-14}$ & 0.45-0.55  & pnQRPA & \cite{qrpa8}  \\
$5.5-6.3 \times 10^{-14}$ & 1.37-1.57 & RQRPA & \cite{qrpa8}  \\
$2.7-3.2\times 10^{-15}$ & 0.07-0.08& SCRQRPA & \cite{qrpa8}   \\
$1.85 \times 10^{-14}$ & 0.46 & pnQRPA & \cite{stoica}  \\ 
$1.21 \times 10^{-14}$ & 0.30  & RQRPA &\cite{stoica}   \\
 $3.63 \times 10^{-14}$ & 0.91  & full-RQRPA &\cite{stoica}  \\
$6.50 \times 10^{-14}$ & 1.62 & SQRPA &\cite{stoica}   \\
$2.88 \times 10^{-13}$ & 7.20 & VAMPIR& \cite{vampir}   \\
 $1.58\times 10^{-13}$ & 3.95 & Shell Model &  \cite{wick}   \\
 $1.90 \times 10^{-14}$ & 0.47 & Shell Model&  \cite{etienne}  \\
\hline
\end{tabular}
\end{table}
\newpage

\end{document}